\documentclass{WileyMSP-template}
\usepackage{graphicx}
\usepackage{dcolumn}
\usepackage{bm}
\usepackage{gensymb}
\usepackage{amsmath,amssymb}
\usepackage{color,soul}
\begin{document}

\newcommand{\cdas}{Cd${}_3$As${}_2$}
\newcommand{\ooz}{[1$\bar{1}$0]}

\pagestyle{fancy}
\rhead{\includegraphics[width=2.5cm]{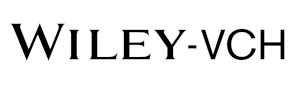}}

\title{Toward Tunable Magnetic Dirac Semimetals: Mn Doping of Cd$_3$As$_2$}

\maketitle


\author{Anthony D. Rice*}
\author{Ian Leahy}
\author{Herve Ness}
\author{Andrew G. Norman}
\author{Karen N. Heinselman}
\author{Chun-Sheng Jiang}
\author{David Graf}
\author{Alexey Suslov}
\author{Stephan Lany}
\author{Mark Van Schilfgaarde}
\author{Kirstin Alberi}

\begin{affiliations}
A.D. Rice, I. Leahy, A.G. Norman, C.-S. Jiang, K.N. Heinselman, M. Van Schilgaarde \\
National Laboratory of the Rockies, Golden, CO 80401, USA\\
Email Address:Anthony.Rice@NREL.gov

H. Ness\\
Department of Physics, King's College  London, Strand, London WC2R 2LS, UK

D. Graf, A. Suslov\\
National High Magnetic Field Laboratory, Tallahassee, FL 32310, USA

K. Alberi\\
National Laboratory of the Rockies, Golden, CO 80401, USA\\
Renewable and Sustainable Energy Institute, University of Colorado Boulder, Boulder, CO, USA

\end{affiliations}


\keywords{Topological Materials, Epitaxy, Magnetism}

\begin{abstract}

Magnetic impurities provide a route toward increasing functionality in electronic materials, often enabling new device concepts and architectures. In the case of topological semimetals, dilute magnetic doping presents a particularly attractive approach for inducing a Dirac to Weyl phase change via time reversal symmetry breaking. However, efforts to realize changes in the electronic structure have been limited by challenges in incorporating magnetic impurities into crystals with sufficiently high electron mobilities to detect them via transport or spectroscopic techniques. Here, we demonstrate incorporation of Mn into Cd$_3$As$_2$ Dirac semimetal thin films grown by molecular beam epitaxy (MBE). Using As-rich growth conditions and [001] oriented thin films, Mn compositions of $>$10\% are achieved. Films contain uniform distributions of Mn with no evidence of secondary phases and exhibit electron mobilities greater than 10,000-30,000 cm${}^2/$Vs up to 5\% Mn. An evolution in the magnetization behavior along with the emergence of a second quantum oscillation frequency at low Mn concentrations provide preliminary evidence of Mn-induced changes in the electronic structure that are consistent with a Weyl phase. This work demonstrates the potential of magnetically doping topological semimetal thin films and a pathway for synthesizing them. 

\end{abstract}

\section{Introduction}
Alloying electronic materials with magnetic impurities has long served as a route to manipulate their properties\cite{Furdyna1986DMSReview}. For example, Mn doping of III-V and II-VI compound semiconductors in the dilute (few atomic percent) range has been pursued for several decades to couple charge and spin transport, introduce ferromagnetism and enhance the ability to modulate optical properties with applied magnetic fields. Combined with thin film epitaxy, magnetic impurity doping has enabled new types of devices, including spin field effect transistors\cite{DattaDas1990} and other spintronic devices\cite{Zutic2004Review}, spin selective contacts\cite{Smith2001SpinContacts}, and spin light emitting diodes\cite{Chang2025SpinLED}. A similar approach promises to unlock additional functionality in newer generations of quantum materials. 

Topological semimetals (TSMs) are one such candidate class of materials\cite{Yan2017}. At their core, time reversal symmetry plays a role in the type of topological phase that is presented. This symmetry is required in the Dirac phase. However, it may be absent in the Weyl phase, implying that it could be possible to switch between the two through the incorporation of magnetic impurities. Cd$_3$As$_2$ is a primary candidate for studying this effect, as it is a Dirac semimetal that can be grown by conventional thin film synthesis methods with high electron mobilities ($>$ $\sim$10$^4$ cm$^{2}$/V-s). Theoretical studies predict substitutional replacement of Cd with a magnetic ion like Mn can break time reversal symmetry, splitting the Dirac points into pairs of Weyl points -- similar in effect to the application of very high external magnetic fields \cite{Ness2025MnCd3As2}. The ability to tune the electronic structure between these phases and the optical and electron transport properties that follow with small composition shifts would create a new pathway to design the properties of topological semimetals without creating an entirely new material. Potential applications include magnetic devices sensitive to changes in their Berry curvature, which has been previously showed to be controlled through strain\cite{Chi2023CrTeStrain}, magnetization\cite{Xue2023TunableAHE}, and symmetry \cite{Li2024TunableHallStates}.

In assessing the prospects for Mn doping Cd$_3$As$_2$, it is instructive to compare it with Mn doping in two related compounds: GaAs and CdTe. Manganese substitutes for the cation in both semiconductors, but it presents subtle, yet important, differences in its behavior. It acts as an isovalent dopant in CdTe, in which Mn has 5 occupied $d$ states far below the Fermi level and 5 unoccupied states far above it\cite{Furdyna1988}. Its behavior in Cd$_3$As$_2$ is predicted to be similar to CdTe in this respect. States at the Fermi level retain \textit{sp} character of the parent compound, with an additional spin splitting of states near the band edges.  Thus, they supply an internal magnetic field as predicted to also occur in Mn-doped Cd$_3$As$_2$~\cite{Ness2025MnCd3As2}. On the other hand, Mn acts as a $p$ type dopant in GaAs\cite{Munekata1989}. The majority Mn-$d$ states are shallow and sit near or at the Fermi level, destroying the itinerant nature of the system. (Mn,Ga)As is ferromagnetic with Zener-like double exchange interactions~\cite{MnGaAsexchange}, while (Mn,Cd)Te is antiferromagnetic, coupled via a superexchange mechanism~\cite{Larson1990}.

In practice, however, Mn also shows a propensity to incorporate in interstitial lattice sites. Interstitial Mn$_i$ acts as an $n$ type dopant in both GaAs and CdTe. It is also well known that Mn$_i$ couples antiferromagnetically to substitutional Mn$_{\mathrm{Ga}}$ in GaAs and is deleterious to device performance\cite{Yu2002Mni}. Cd$_3$As$_2$ may be thought of as an anti-fluorite structure with ordered Cd vacancies. Mn$_i$ could have a formation energy comparable to the Mn$_\mathrm{Cd}$ in this structure, particularly when electrically compensated by other charged defects.  The relative stability of Mn$_i$ and Mn$_\mathrm{Cd}$ and their impact on the electronic structure are important considerations for realizing Weyl systems in the dilute Mn doping regime.

Incorporation of magnetic elements into thin films has also remained a significant challenge to dilute magnetic element doping strategies in application-relevant thin films. Low growth temperatures are typically used to counteract its relatively low solubility and formation of secondary phases (e.g. MnAs) in (Mn,Ga)As\cite{Edmongs2007GaMnAsPhases}. Growth of Mn-doped topological insulators with homogeneous Mn incorporation presents similar challenges, with a tendency for Mn to segregate towards the surface in Bi$_2$Se$_3$\cite{2012ZhangMnBiSe}. More recently, efforts have been made to incorporate Mn into Cd$_3$As$_2$ thin films grown by molecular beam epitaxy (MBE) \cite{2022PRM_MnCdAs}. Despite the low temperatures already used for Cd$_3$As$_2$ growth, Mn preferentially behaved as a surfactant, forming a Mn-rich surface layer instead of incorporating into the film. The other report of Mn-doped Cd$_3$As$_2$ thin films synthesized by other methods presented films with very low electron mobilities (\textless 1000 cm$^2$/V-s \cite{Wang_2020MnCd3As2ThinFilms}) compared to high quality Cd$_3$As$_2$ thin films, calling into question how Mn incorporation impacts the microstructure. Approaches that overcome these limitations to produce Mn-doped alloys that retain the high electron mobilities of Cd$_3$As$_2$ would open the door to experimentally verifying the predicted topological phase change and exploring many new device architectures. 

Here, we show that Mn can be uniformly incorporated into Cd$_3$As$_2$ thin films grown by MBE by carefully tuning the growth window. Arsenic-rich flux ratios and growth on the higher energy (001) surface are key factors in aiding Mn incorporation into a smooth film. Our single crystal, structurally phase-pure films, with Mn concentrations spanning from $<$1 to $>$10 atomic percent Mn, additionally exhibit high electron mobilities in the 10$^3$ – 10$^4$ cm$^2$/V-s range, demonstrating the ability to achieve both high Mn concentrations and high electron mobilities. As with GaAs and CdTe, magnetization measurements and structural characterization indicate that some Mn may incorporate on interstitial sites. Interestingly, magnetotransport measurements provide preliminary evidence that Mn, with its moments ferromagnetically aligned by a small applied magnetic field, may indeed alter the electronic structure by splitting the Weyl nodes. These advances lay the foundation for designing and synthesizing dilute magnetically-doped topological semimetals wherein their electronic structure can be dynamically tuned.

\section{Results}

\subsection{Enhanced Mn Incorporation Via Tailored Growth Conditions}

Incorporating Mn into covalently bonded crystals, such as semiconductors and topological insulators or semimetals, has been notoriously difficult because of its tendency to surface segregate. Segregation is often aided by its high surface mobility as well as strain and electronic effects. Typical strategies to avoid this surfactant behavior include lowering the growth temperature to kinetically limit the exchange mechanisms that allow the atoms to remain on the surface during growth. For example, Mn-doping of GaAs is carried out at substrate temperatures more than 300 $\degree$C lower than those used for undoped GaAs to promote incorporation as well as prevent secondary phase formation.\cite{1998OhnoMnAs}. Because Cd$_3$As$_2$ is already grown at such low temperatures ($\sim$115 $\degree$C), alternative strategies must be used to enhance incorporation. Here, we tailor the growth conditions in two ways to drive uniform Mn doping throughout a smooth film: 1) using As-rich conditions to reduce the competition of Mn with Cd for bonding with a limited number of As$_4$ molecules adsorbed to the surface, and 2) growing on a higher energy (001) surface instead of the lower energy [112] surface to reduce surface roughness. \\

\begin{figure}
  \includegraphics[width=\linewidth]{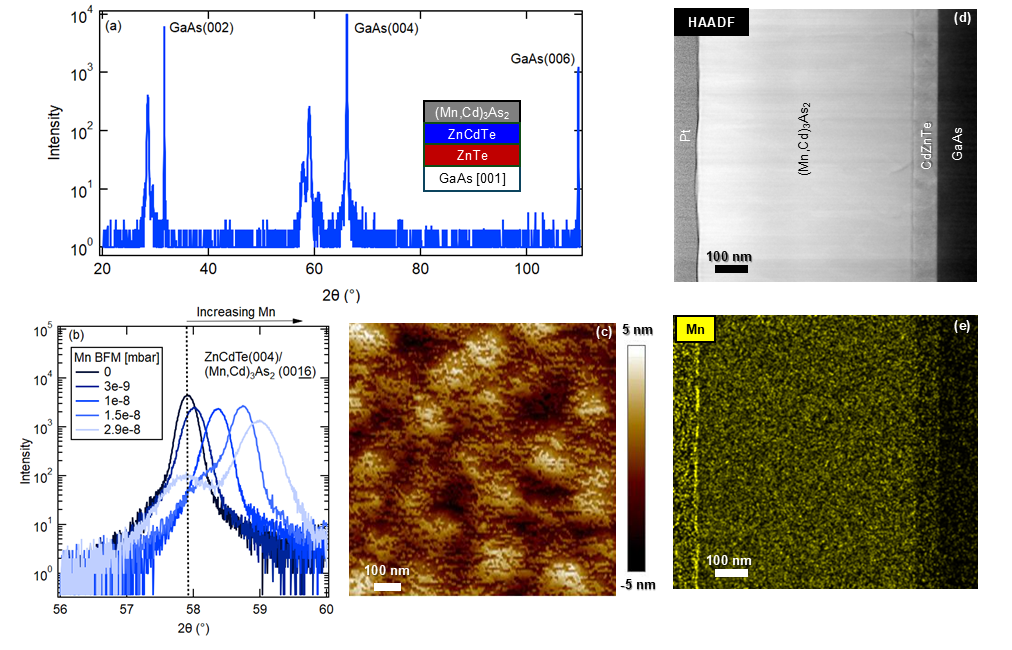}
  \caption{Summary of thin film growth a) Survey XRD scan of a (Mn$_{0.05}$Cd$_{0.95}$)$_3$As$_2$ and b) XRD scans around the (00\underline{16}) (Mn,Cd)$_3$As$_2$ peak as a function of Mn BFM flux. c) AFM of a 1x1 $\mu$m area and (d-e) STEM HAADF image and Mn EDS elemental map of the film stack. Films are smooth, with a gradual shift in lattice parameter in the (Mn,Cd)$_3$As$_2$ peaks, no visible secondary peaks, and uniform Mn content through the bulk of the layer. For higher Mn content, the ZnCdTe(004) buffer layer peak may be visible near 58$\degree$.}
  \label{fig1}
\end{figure}

Initial growth attempts using [112] oriented films showed evidence of Mn incorporation, but given the low energy of this surface, and much lower temperatures than Mn-alloys are typically synthesized, films contained extremely large surface features (Root Mean Square (Rq) surface roughness $>$25 nm, see supporting information note 1). We hypothesize this is due to the small energy cost of increasing the surface area combined with poor Mn surface mobility at 115 $\degree$C.  
We therefore utilize a higher energy (001) surface to remove these features by adding an energy barrier to increasing roughness. 

The previous reported attempt to synthesize (Mn,Cd)$_3$As$_2$ via MBE resulted in films that exhibited no shift in the main XRD peak associated with (Mn,Cd)$_3$As$_2$ but instead the emergence of secondary peaks attributed to a surface Mn-layer, which was confirmed using TEM\cite{2022PRM_MnCdAs}. Those growths were performed under Cd-rich conditions. However it's been demonstrated that a wide range of As/Cd ratios may result in single phase growth Cd$_3$As$_2$, including As-rich, which also influences the point defect density due the associated differences in chemical potentials \cite{Nelson2023}. Utilizing As-rich growth conditions should also reduce the competition between Mn and Cd for bonding with a limited number of As$_4$ molecules are the surface and prevent the scenario wherein Cd preferentially incorporates in the growing crystal over Mn. 

Figure \ref{fig1} summarizes the structural and compositional properties of our (Mn,Cd)$_3$As$_2$ films grown under As-rich growth conditions and on (001) surfaces. A single set of diffraction peaks is observed in the x-ray diffraction scans for each layer (see figure \ref{fig1}a), with no evidence of additional Mn-rich  phases. With increasing Mn flux, a gradual shift toward smaller lattice parameters is also observed in the (Mn,Cd)$_3$As$_2$ peaks (see figure 1b), suggesting systematic incorporation of Mn into the crystal. Table \ref{MnFluxTable} summarizes the relationship between Mn flux, calculated Mn composition, and (00\underline{16}) peak positions. Compositions are calculated using Rutherford backscattering spectrometry (RBS) standards of Mn grown on Si to correlate Mn beam flux monitor (BFM) readings with Mn atoms/s-cm$^2$, growth time, and overall film thickness. Fluxes resulting in concentrations of approximately 15\% Mn have been employed to date, with no evidence of secondary phase formation. While peaks are similarly narrow below 10$\%$ Mn, peak widening is observed (via full- width half-max (FWHM)) along with smaller relative shifts (i.e. non-linear) in the (00\underline{16}) peak position at higher Mn compositions. This widening and reduced relative shrinking of the lattice could be explained by fewer substituional Mn and more interstitial Mn. Atomic force microscopy (AFM) scans (see figure \ref{fig1}c), reveal Rq rougness values of $\sim$1.5 nm, even across a 10x10 $\mu$m area, comparable to morphology in undoped Cd$_3$As$_2$[001]\cite{Rice2022AFM}. No additional evidence of surface clustering of Mn was observed.

\begin{table}
\begin{center}
\begin{tabular}{ |c|c|c|c|}  
\hline
 Mn Flux (mbar) & Mn $\%$ & (00\underline{16}) Peak Position ($\degree$) & FWHM ($\degree$)  \\ [0.5ex] 
 \hline
 \hline
 3.0e-9 & 1.9 & 57.99 & 0.291\\ 
 1.0e-8 & 5.9 & 58.37 & 0.281 \\ 
 1.4e-8 & 8.1 & 58.72 & 0.270 \\ 
 2.9e-8 & 15 & 58.90 & 0.411 \\ 
 \hline
 \end{tabular}
\caption{Summary of relationship between measured flux, calculated Mn concentration, (00\underline{16}) peak position and FWHM.}
\label{MnFluxTable}
\end{center}
\end{table}

\begin{figure}
\centering
  \includegraphics[width=0.7\linewidth]{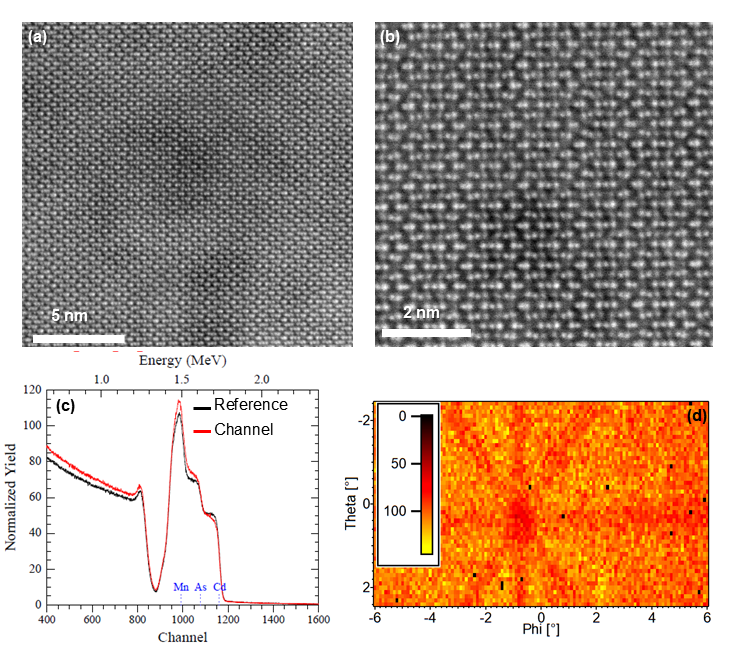}
  \caption{Investigations of short range ordering of Mn. a) Cross-sectional high resolution STEM HAADF image near layer surface. b) Higher magnification STEM HAADF image. c) Comparison of a ``random" (non-channeling) and channeling axis spectra and d) RBS channeling crystal image of the Cd signal from a $8\%$ Mn film. Higher intensity indicates higher backscattered yield. }
  \label{fig2}
\end{figure}

The uniformity of Mn incorporation was examined via scanning transmission electron microscopy (STEM) and energy dispersive x-ray spectroscopy (EDS). A STEM high-angle annular dark field (HAADF) image of a (Mn$_{0.06}$Cd$_{0.94}$)$_3$As$_2$ thin film is shown in figure 1d as an example. No secondary phases or inclusions were detected. Within the resolution of this technique, a uniform distribution of Mn is observed throughout the layer, with some evidence of a very thin (5 nm, $\sim$1$\%$ of the total layer thickness) Mn-rich surface. Higher resolution EDS data suggests this layer is amorphous and has a Mn composition approximately twice that of the bulk layer (see supporting information note 2). These results suggests that unlike in the previous report\cite{2022PRM_MnCdAs}, where Mn remained on the surface, the excess As forces incorporation throughout the film. 

Beyond confirming incorporation of Mn throughout the film, previous observations in MBE-grown (Mn,Ga)As suggest localized clustering could be an issue \cite{RAEBIGER2005MnGaAsClustering,Moreno2002MnGaAsNanoclusters}. Figure \ref{fig2} summarizes further characterization to examine the microstructure of Mn. High resolution STEM HAADF imaging was performed in an aberration corrected STEM, and representative images at different magnifications obtained from the  (Mn$_{0.06}$Cd$_{0.94}$)$_3$As$_2$ film are shown in figure 2a and b. No secondary phases such as MnAs, or obvious clusters or long range ordering of the Mn atoms are visible. RBS spectra measured along the channeling axis (red) and ``random" orientation away from the channeling axis (black) are displayed in Figure 2c. Figure \ref{fig2}d shows the channeling signal from Cd, which confirms the crystallinity of the sample, with a  channeling axis and a faint start pattern around the middle. The channeling crystal image, along with the channeling and ``random" (non-channeling) scans, indicate a significant amount of dechanneling, in part due to defects in the sample. This is supported by the channeling minimum yield is around $\chi_{min} \approx$ 80\% (backscattering yield ratio). It is important to note here that several defect types may contribute to dechanneling, including point defects and dislocations. Previous studies of our Cd$_3$As$_2$ epitaxial films have revealed dislocation densities as high as $10^{8}\ \mathrm{cm}^{-2}$\cite{Rice2019}. 


\subsection{Magnetization}

Magnetization and magnetotransport measurements provide further insight into how Mn atoms are locally positioned in the lattice, how they interact, how they influence electron transport, and potential signatures of how the topological phase may evolve with Mn concentration. We first examine the magnetization behavior to understand how the Mn atoms interact within the lattice. Figure \ref{figmag} summarizes the in-plane magnetization measured for (Mn,Cd)$_3$As$_2$ with increasing Mn concentration at 2 K. For $0.3 \%$ Mn, the magnetization is paramagnetic-like, reaching a moment near 2.5 $\mu_B/$Mn by 1 T. As the Mn concentration increases, the saturation field scale and the saturated moment decrease, reducing to $0.3\,\mu_B/\textrm{Mn}$ for $3.1\%$ Mn. Near $8\%$ Mn, the moment per Mn reaches a minimum. A clear open hysteresis loop is only observed for 15.6$\%$ Mn, though the saturation moment remains small. 

The reduction of the moment per Mn below the expected free Mn moment ($\sim$ 5.92 $\mu_B/$Mn) is indicative of antiparallel alignments between a fraction of close lying Mn$_\mathrm{Cd}$ and/or Mn$_{i}$ similar to (Mn,Ga)As. We hypothesize that the rapid reduction of the saturation moment with increasing Mn concentration reflects an imperfect but increasing concentration of antiparallel couplings and ferrimagnetic-like behavior, as schematically depicted in figure \ref{figmag}a. Low Mn doping ($<1.9\%$) likely results in large distances and weak couplings between neighboring Mn atoms, preserving a large net moment per Mn ion. At intermediate doping ($1.9-4\%$), the probability of finding two or more Mn ions sitting in close proximity increases (e.g. on neighboring substitutional and interstitial sites \cite{Kulatov2021}). The experimentally measured reduction in the moment per Mn at these intermediate concentrations signals some antiferromagnetic coupling. Around $8\%$ Mn doping, the majority of Mn atoms are maximally antiferomagnetically coupled to partner Mn atoms, as reflected in the small net moment. With further increasing doping, the increased moment and small hysteresis loop likely results from competition or frustration between higher concentrations of Mn ions. 

\subsection{Electrical Transport}
To study the effects of Mn doping on the electronic structure and transport properties, we turn our attention to the 2 K magnetotransport behavior. Figure. \ref{figtransport}a,b show the fractional magnetoresistance ($\mathrm{FMR}\equiv (\rho_{xx}(H)-\rho_{xx}(0 \mathrm{T}))/\rho_{xx}(0 \mathrm{T})$ and Hall resistivity for samples up to $8.3\%$ Mn and fields up to 31 T. We use the low field slope of $\rho_{xy}$ ($<1$ T) to extract the electron density (figure \ref{figtransport}c) and the zero field longitudinal resistivity, $\rho_{xx}(0\mathrm{T})$, to calculate the mobility (figure \ref{figtransport}d). The spread in electron densities is similar to what is observed in undoped films ($1-10\times 10^{17}$ cm${}^{-3}$). The addition of Mn does not substantially or systematically alter the electron density at these concentrations, consistent with isovalent substitution. The mobility of undoped Cd$_3$As$_2$[001] bulk-like films ($>100$ nm\cite{Goyal2018ThicknessDependence}) is typically between 10,000-30,000 cm${}^2/$Vs, depending on sample quality and measurement temperature. The highest mobility sample in the series reaches a notably high value for a Mn-doped Cd$_3$As$_2$[001] film with 0.5$\%$ Mn concentration at 50,450 cm${}^2/$Vs at 2K, potentially enabled by Mn surfactant-related effects on defect formation\cite{2022PRM_MnCdAs}. At higher concentrations, adding more Mn decreases the mobility, consistent with typical alloy scattering.

Quantum oscillations of $\rho_{xx}$ and $\rho_{xy}$ may also offer signatures of a change in electronic structure and topological phase. Samples across all Mn concentrations show clear quantum oscillations. Focusing on $\rho_{xy}$ for the $5.4\%$ and $8.3\%$ Mn samples, we note the peculiar shape of the Hall resistivity. This step-like behavior superimposed on a linear background has recently been observed in Cd$_3$As$_2$ films and other low electron density semimetals, originating from the quasi-quantum Hall effect \cite{Leahy2025,Galeski2021}. Analyzing quantum oscillations in Cd$_3$As$_2$ films is challenging: low electron densities result in low frequency oscillations while film mobilities restrict the field ranges where clear oscillations are visible. At the same time, oscillations are superimposed on a large, nonsaturating magnetoresistance background originating from scattering from charged impurities\cite{Nelson2023,skinner_coulomb_2014}. As a result, oscillations must be carefully separated from the magnetoresistive background and few oscillations are visible. When the harmonic content of oscillations is focused at low frequencies ($<50$ T), residuals introduced by simple polynomial fitting obscure the resulting fast Fourier transform (FFT) spectra. To minimize these effects, we extract the oscillatory component of $\rho_{xx}$ by taking derivatives, restrict the FFT analysis range to where oscillations are visible in $\partial_H^2\rho_{xx}$ vs. inverse field,  fit and remove a simple polynomial background to $\partial_H^2\rho_{xx}$ in inverse field, and finally perform the FFT.

Figure \ref{figtransport}e shows the FFT spectra for samples up to $8.3\%$ Mn for $H\parallel I$. We analyze oscillations in this orientation to minimize the contributions to the magnetoresistance from scattering from charged disorder. Arrows in the figure denote bulk oscillation frequencies. We attribute the small shoulders at higher frequencies above the main peak to higher harmonics. With just $0.5\%$ Mn, the main oscillation peak splits into two peaks. For 1.9$\%$ Mn, the second peak is even more distinct. Above $3.1\%$, the bulk peak appears broadened and it is difficult to distinguish two distinct bulk peaks, coincident with the shrinking sample magnetization. 

\begin{figure}
  \centering
  \includegraphics[width=0.5\linewidth]{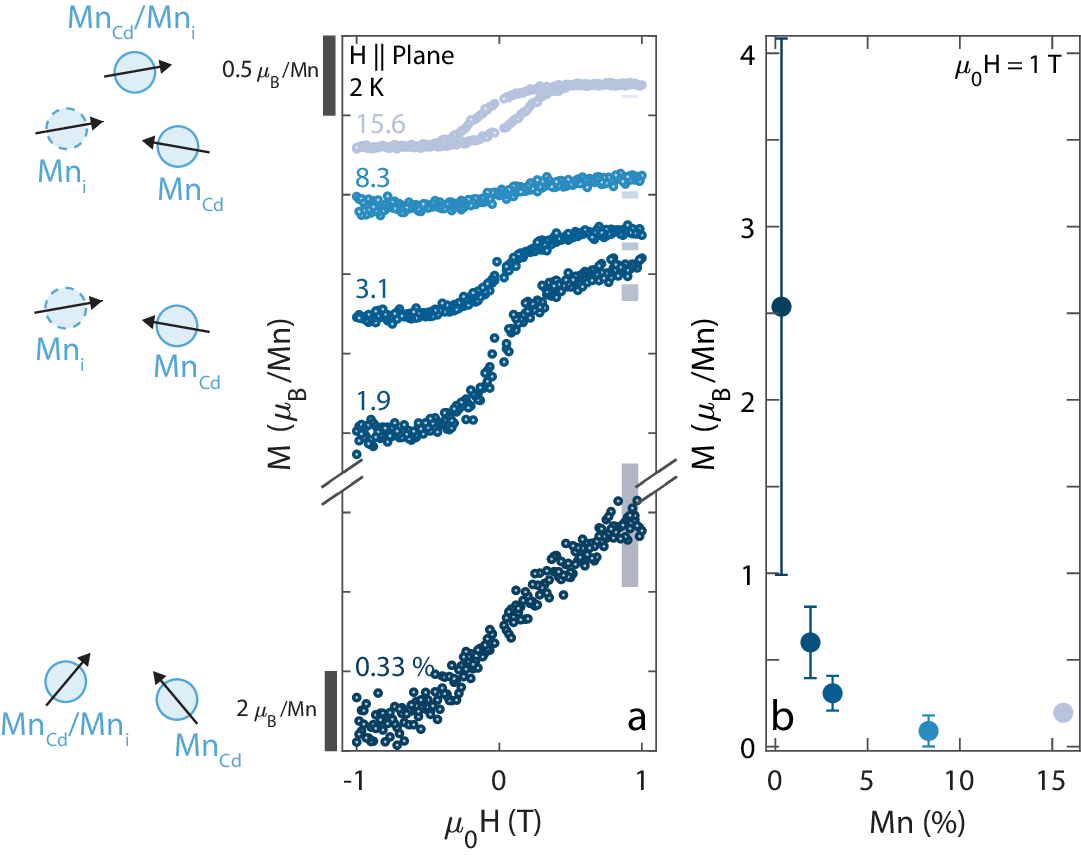}
  \caption{Magnetization of (Mn,Cd)$_3$As$_2$ films with varying Mn concentrations at 2 K. A linear diamagnetic background from the substrate has been removed. a) Magnetization vs. Field loops for films with different Mn concentrations. A clear hysteresis loop is only observed at high Mn concentrations. The vertical scale is broken and loops are offset for clarity. Light colored boxes represent error in the magnetization measurement. b) Value of the magnetization at 1 T vs. Mn concentration. At low Mn concentrations, the moment per Mn atom is largest. The moment per Mn rapidly decreases with increasing Mn concentration. }
  \label{figmag}
\end{figure}
\begin{figure}
  \centering
  \includegraphics[width=0.85\linewidth]{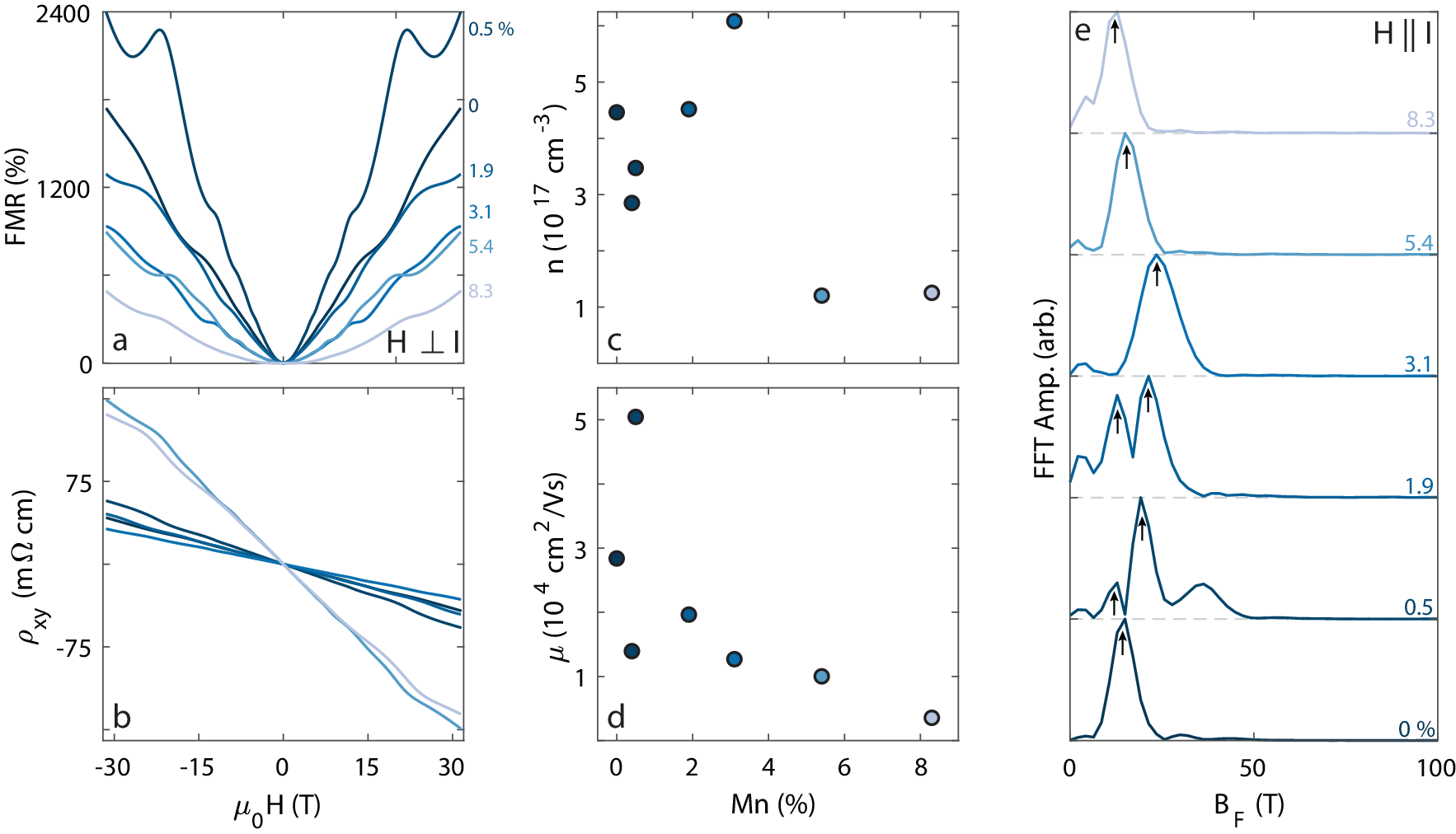}
  \caption{2 K a) Fractional magnetoresistance and b) Hall resistivity for (Mn,Cd)$_3$As$_2$ films at 2 K for $H\perp I$. c) Electron density and d) Mobility as a function of Mn percentage. e) Fourier transforms of quantum oscillations extracted from $\partial_H^2\rho_{xx}$ for $H\parallel I$. Arrows denote the bulk oscillations which split at low Mn doping. }
  \label{figtransport}
\end{figure}

In a previous theoretical work, we have shown that the addition of Mn lifts the degeneracy of the bands near the Dirac point, transforming the Fermi surface into sets of nested ellipsoidal pockets of differing size \cite{Ness2025MnCd3As2}. We have also shown that applied magnetic fields emulate the effects of Mn doping: an external field giving a 72 meV Zeeman energy applied to pristine Cd$_3$As$_2$ reproduced the band structure effects of just $2\%$ Mn doping when all Mn moments were ferromagnetically aligned. We hypothesize here that partial ferromagnetic alignment of Mn ions in these films splits the Dirac points, producing two nested Fermi surfaces that, in turn, contribute the two oscillation frequencies. Using our previously developed theoretical framework, we calculate that a Zeeman energy of 22-29 meV is needed to reproduce the ellipsoid surfaces consistent with the measured split oscillation frequencies. This would theoretically correspond to ferromagnetic alignment of Mn ions carrying $1.8$-$2.4$ $\mu_B$ each, in agreement with figure \ref{figmag}b (see Supporting Information Note 3 for details).


From the magnetization behavior, it is clear that magnetic doping of Cd$_3$As$_2$ must proceed with care. Magnetic doping of Dirac semimetals is of interest for lifting the nodal point degeneracy, transmuting the system to a Weyl semimetal. With Mn dopants, it has been theoretically shown that low Mn concentrations can lift this degeneracy when ferromagnetically coupled\cite{Ness2025MnCd3As2,KULATOV2021MnCd3As2}. In the limit where Mn ions are not coupled together, a small applied field can orient the Mn moments, acting as an amplifier of applied magnetic fields that affects the band structure and Dirac point splitting. In the experimental reality, higher Mn concentrations do not lead to larger saturation moments because of greater degreens of antiferromagnetic Mn-Mn coupling. 




\section{Discussion}

Magnetic doping of topological semimetals opens new doorways for novel device functionalities anchored to the manipulation of topological phases. We have demonstrated a pathway to Mn-doping Cd${}_3$As${}_2$ thin films with high enough quality to start to probe the resulting change in the electronic structure. We find relatively uniform Mn incorporation throughout the film and an absence of the noticeable clustering observed in magnetic doping of some semiconducting systems. Enabled by this growth, we are able to explore questions about the realistic effects of magnetic doping in Cd${}_3$As${}_2$.

Our primary results related to the behavior of our (Mn,Cd)$_3$As$_2$ films are as follows. We observe an in-plane magnetization originating from the alignment of Mn magnetic moments under the application of a small external magnetic field. The magnetic moment per Mn ion is closest to the free Mn ion moment at low Mn concentrations and decreases with added Mn. At the same time, we detect a splitting  of the main bulk quantum oscillation frequency consistent with the magnetization and the anticipated effect of Mn to break time reversal symmetry \cite{Ness2025MnCd3As2}. 

Many open questions remain about the origins of this behavior. The first is, what happens to the electronic structure? Theoretical predictions suggest the incorporation of Mn transitions Cd$_3$As$_2$ from a Dirac to Weyl semimetal. We have shown that the addition of Mn clearly alters electronic properties, but experimentally verifying this topological phase transition presents challenges. Angle-resolved photoemission spectroscopy (ARPES) would provide the most definitive signatures of a Weyl phase, either through observation of band structures similar to this prediction, or through the observation of open Fermi arcs. The high vapor pressure of Cd$_3$As$_2$ prevents straightforward transport of samples to beamlines, instead requiring vacuum suitcases or in-situ growth.  Signatures in transport commonly associated with the Weyl phase, such as negative linear magnetoresistance from the chiral anomaly, or the anomalous Hall effect, could also arise from other factors, such as current jetting\cite{Arnold2016CurrentJetting}, or conversely, could not be present due to other factors, such as crystalline disorder\cite{Chen2015WeylDisorder,Su2017WeylDisorder, Zheng2016WeylScattering}. Instead, we rely on the observed peak splitting in quantum oscillations as a signature consistent with the existence of separated Weyl nodes, but it is not definitive on its own. Non-reciprocal optical responses are also indicative of broken symmetries, but again, are not always indicative nor a requirement of Weyl semimetals \cite{Kotov2018NonReciprocity}.

A second open question from the results presented here is ``what fraction of the of the Mn atoms reside in interstitial sites?" Preliminary results of first-principles calculations of Mn$_\mathrm{Cd}$ and Mn$_i$ incorporation using an analogous approach as in Refs.\cite{brooks2023band, brooks2025pseudo} for intrinsic defects and extrinsic dopants suggests that 
Mn$_\mathrm{Cd}$ substitution has low formation energies in the range between 0.3 - 0.5 eV, depending on the three non-equivalent Cd sites and the Cd- vs As-rich growth conditions. The Mn$_i$ acts as a double donor whose formation energy depends on $E_\mathrm{F}$, similar to the intrinsic Cd interstitial. The energy cost of transforming a substitutional atom into vacancy-interstitial pair is lower for Mn than for Cd, indicating an increased tendency for disordering on the underlying anti-fluorite lattice due to Mn incorporation. 

Experimentally, the deviation between the magnetization that is measured in our films and expected values suggest that Mn dopants appear to be coupling antiferromagnetically, reducing the net moment per Mn. Antiferromagnetic coupling has been associated with Mn$_i$ in (Mn,Ga)As, but coupling is also possible between adjacent Mn$_\mathrm{Cd}$ as well. Likewise, the high degree of dechanneling in the RBS spectra is consistent with high defect concentrations, but Mn$_i$ are not the only possible sources of dechanneling. On the other hand, the electron concentration in the Mn-doped films does not strongly increase relative to undoped Cd$_3$As$_2$ films, which would otherwise occur for high concentrations of un-compensated Mn$_i$ defects. The electron mobility also remains high through several percent Mn, suggesting that Coulomb scattering from charged Mn$_i$ defects does not change excessively as dilute concentrations of Mn are added. A take-away from this study is that Mn$_i$ defects likely play a role in the magnetization and electronic structure of Mn-doped Cd$_3$As$_2$, but much more focused and thorough studies are needed to precisely determine their concentrations.

A follow-on question to ask is ``how can Mn$_i$ defects be controlled or reduced?" We anticipate that the magnetic behavior could be enhanced further at intermediate Mn values by reducing the Mn$_i$ concentration. Mn$_i$ have been slowly annealed out at low temperatures in (Ga,Mn)As, with interstitials migrating to the surface~\cite{Edmonds2004}, \cite{sadowski2005solidphaseepitaxyferromagnetic}. However, this process is best suited for very thin films and is not guaranteed to work in Cd$_3$As$_2$. Re-adjusting buffer layer composition to match the smaller Mn-doped structure, or controlling compressive strain, may also produce different Mn site distributions if there is a noticeable difference in the equilibrium lattice constant based on Mn site position. Finally, previous changes were observed in the magnetism of (Ga,Mn)As thin films based on the hole density and were explained with a Zener model\cite{Dietl2000MnGaAsZener}. Similar studies could be performed combined extrinsic electron doping with group VI elements \cite{Rice2023} to see how these properties evolve and give further insight into the mechanism, or the electronic structure. 

Beyond site disorder, ways to further improve overall film quality could prove fruitful. The film mobility remains high through the solid solution shown in figure \ref{figtransport}, suggesting the addition Mn itself does not add a potential that readily scatters electrons. 
An additional surfactant could improve the crystal quality without producing surface segregation of Mn, such as Bi\cite{TIXIER2003BiSurfactant}. Substrate temperatures are limited because of the high vapor pressure of Cd, As, and Cd$_3$As$_2$, but capping and post growth anneal may prove useful as well\cite{nakazawa2018structural}.  

\section{Conclusion}

Incorporation of a magnetic dopant, in this case Mn, has been demonstrated in Dirac semi-metal Cd$_3$As$_2$. Through the use of [001] oriented films grown under As-rich conditions, uniform Mn incorporation in excess of $15\%$ are demonstrated while retaining high electron mobilities. Magnetization measurements provide a picture of the evolving nature of Mn magnetic moment alignment, while magnetoresistance measurements provide preliminary evidence of a change in the electronic structure. Importantly, a second oscillation frequency in $\rho_{xy}$ is observed and is consistent with splitting of Dirac cones due to breaking time reversal symmetry. More work is necessary to confirm this initial finding. Overall, these results establish a path toward manipulating the topological phase of TSMs using dilute magnetic doping along with small applied magnetic fields.

\section{Methods}
Thin films were grown on GaAs[111]B and GaAs[001] substrates by MBE. Substrates were miscut 2$\degree$ toward [112] and 4$\degree$ toward [110]A respectively. A 300nm GaAs buffer was grown at 580 C with elemental Ga and As$_2$ fluxes, and substrates were then As capped before being loaded into a second chamber. There, a ZnTe/Zn$_x$Cd$_{1-x}$Te buffer structure was grown, as described elsewhere, to serve as a lattice-matched template for (Mn,Cd)$_3$As$_2$ film growth.\cite{Rice2019}\cite{Rice2022AFM} (Mn,Cd)$_3$As$_2$ films were then grown at 115$\degree$C using an As$_4$ overpressure of 3-5/1 with a growth rate of approximately 500nm/hr. Mn was supplied from a standard Knudsen-cell (K-cell) with a variable flux. All data shown comes from films with a thickness of 500-700nm. Mn composition is approximated using RBS standards, the measured Mn cell flux, and the film thickness. Scanning transmission electron microscopy (STEM) samples were prepared using a standard lift out method performed in a Ga+ ion focused ion beam (FIB) work station. Ga+ ion FIB sample damage was subsequently removed by low angle, low voltage Ar+ ion milling with the sample cooled using liquid nitrogen. Thin samples were then examined in a Thermo Fisher Spectra 200 aberration corrected STEM operated at 200 kV using HAADF imaging and EDS analysis. Rutherford Backscattering Spectrometry channeling was performed using a 168$\degree$ backscattering geometry. The source was a 2 MeV He+ beam using a model 3S-MR10 RBS system from National Electrostatics Corporation. 

Magnetization measurements were performed using the ACMS II option in the Quantum Design Physical Property Measurement System. Films were cut into $\approx$ 3 x 5 mm pieces, dipped briefly in etchant to remove any excess-Mn on the surface, and mounted to a quartz paddle with a small amount of thinned GE varnish. A simple, linear diamagnetic background is removed from the magnetization curves originating from the GaAs substrate. The total film volume which is used to calculate the moment per Mn atom is estimated using microscopy and ellipsometry.  

(Mn,Cd)$_3$As$_2$ 6-contact Hall bars with electroplated Au contacts were fabricated using standard wet photolithography. A brief dip in etchant is performed prior to fabrication to ensure the removal of any potential Mn-rich surface layer. Electrical transport measurements of the longitudinal and Hall resistivity were simultaneously performed using a low-frequency ($<$ 50 Hz) AC technique with two Stanford Research Systems SRS 860 lock-in amplifiers locked to a Keithley 6221 AC/DC current source in a 31.5 T resistive magnet at NHMFL. Electron densities are extracted from the low field slopes of $\rho_{xy}$.
\medskip

\textbf{Acknowledgements} \par 
This work was authored by the National Laboratory of the Rockies for the U.S. Department of Energy (DOE), operated under Contract No. DE-AC36-08GO28308. Research was performed under the Disorder in Topological Semimetals project funded by the U.S. Department of Energy Office of Science, Basic Energy Sciences, Physical Behavior of Materials program. The views expressed in the article do not necessarily represent the views of the DOE or the U.S. Government. The U.S. Government retains and the publisher, by accepting the article for publication, acknowledges that the U.S. Government retains a nonexclusive, paid-up, irrevocable, worldwide license to publish or reproduce the published form of this work, or allow others to do so, for U.S. Magnetotransport measurements at high magnetic fields was performed at the National High Magnetic Field Laboratory and was supported by the National Science Foundation through NSF/DMR-2128556 and the State of Florida. Government purposes. We acknowledge Patrick Walker for the FIB preparation of STEM samples.
\medskip

%
\bibliographystyle{MSP}

\providecommand{\noopsort}[1]{}\providecommand{\singleletter}[1]{#1}%

\end{document}



\title{Supplementary Information: Toward Tunable Magnetic Dirac Semimetals: Mn Doping of Cd$_3$As$_2$}

\author{Anthony D. Rice}
\email[]{To whom all correspondence should be addressed. Anthony.rice@nrel.gov}
\author{Ian Leahy}
\author{Andrew Norman}
\author{Chun-Sheng Jiang}
\author{Stephan Lany}
\author{Mark Van Schilfgaarde}
\affiliation{National Laboratory of the Rockies, Golden, Colorado 80401, USA}

\author{Herve Ness}
\affiliation{Department of Physics, King's College  London, Strand, London WC2R 2LS, UK}

\author{David Graf}
\author{Alexey Suslov}
\affiliation{National High Magnetic Field Laboratory, Tallahassee, FL 32310, USA}

\author{Kirstin Alberi}
\affiliation{National Laboratory of the Rockies, Golden, Colorado 80401, USA}
\affiliation{Renewable and Sustainable Energy Institute, National Renewable Energy Laboratory and University of Colorado, Boulder, 80301, CO, United States}

\date{\today}

\maketitle

\section{\label{sec:level1}Supplemental Note 1: Growth of (Mn,Cd)$_3$As$_2$(112)}
\begin{figure}[hbt!]
\includegraphics[width=0.5\linewidth]{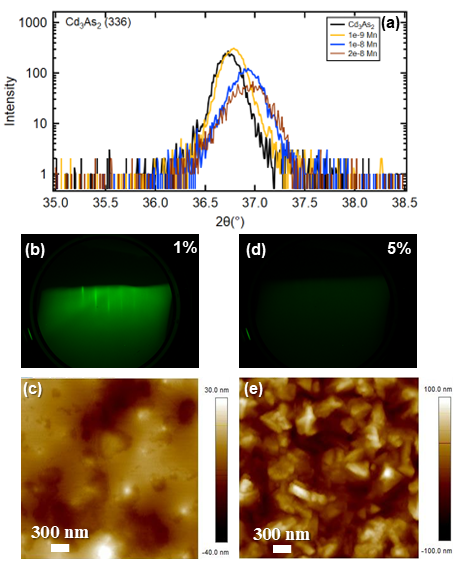}
\caption{\label{figSI1} a) XRD of Cd$_3$As$_2$ (336) peak} b,c) RHEED and AFM (3x3 $\mu$m) of sample with 1$\%$ Mn and d,e) RHEED and AFM of sample with $5\%$ Mn. 
\end{figure}

Initially, growths were performed on GaAs[111] substrates, as described elsewhere \cite{Rice2019}. Figure ~\ref{figSI1} summarizes the results. X-ray diffraction revealed a single set of peaks for each layer in the structure, with a shift in the (Mn,Cd)$_3$As$_2$ peaks toward smaller lattice parameters with increasing Mn flux. Reflection high energy electron diffraction (RHEED) patterns were bright and similar to undoped Cd$_3$As$_2$ films, but increasing Mn content resulted in hazier backgrounds and diminishing pattern intensity. Atomic force microscopy (AFM) revealed films much rougher than normal, with root mean square roughness (Rq) in excess of 25nm for films with $>$5 \% Mn. Because this is the low energy surface, there is little barrier to increasing the surface area, unlike in the (001) face, where the morphology is largely unchanged with increasing Mn. This roughness likely comes from the low growth temperatures (115 $\degree$C), with many Mn compounds optimally grown at 300 $\degree$C or above\cite{2001GaAsMn_MEE}. Points are visible in Fig. \ref{figSI1}c throughout the surface that may be precursors to the larger features seen in Fig. \ref{figSI1}e, possibly due to poor Mn adatom mobility at these temperatures. This is the same surface orientation attempted in the reported negative result, \cite{2022PRM_MnCdAs} and likely also contribute to difficulties in Mn incorporation. 

\newpage

\section{\label{sec:level1}Supplemental Note 2: Amorphous Mn-rich surface layer}

\begin{figure}[hbt!]
\includegraphics[width=0.3\linewidth]{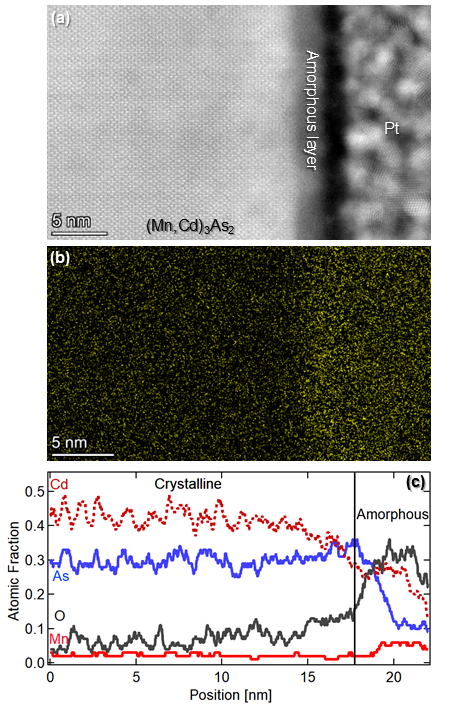}
\caption{\label{figSI2} a) HAADF-STEM image, b) EDS elemental map of Mn, and c) various elemental EDS intensities as a function of depth. All data taken near the surface of a 500nm thick, 5$\%$ Mn film.} 
\end{figure}

The previous report of Mn-doping Cd$_3$As$_2$ revealed that rather than incorporate into the bulk film, the vast majority of Mn was contained in a surface layer $\sim$1/3 of the expected Cd$_3$As$_2$ layer thickness\cite{2022PRM_MnCdAs}. While the data in figure 1 eliminates this extreme case, showing significant Mn incorporation in the film, there is some evidence of a small surface Mn-rich layer. Figure \ref{figSI2} summarizes high resolution scanning transmission electron microscopy (STEM) and energy dispersive spectroscopy (EDS) data taken near the surface, below a post-growth deposited Pt layer used in the sample preparation process. In the case of a $\sim$500nm thick film with 5\% Mn, a 2-3nm thick non-crystalline region is observed at the surface, with a Mn composition of approximately twice that of the bulk ($\sim$10\%). RHEED patterns remain visible throughout growth, suggesting some of this non-crystallinity is likely due to the formation of a native oxide following removal from the growth chamber, since a layer this thick would result in major disruptions to the pattern. 

\newpage

\section{Supplemental Note 3: Quantum Oscillation Splitting }
The inclusion of Mn dopant has been predicted to break time reversal symmetry moving Cd$_3$As$_2$ from a Dirac to Weyl semimetal. Recently, we have theoretically studied the effects of Mn doping on the electronic structure of Cd$_3$As$_2$ \cite{Ness2025MnCd3As2}.  In that work, we have calculated the QSGW band structure for Mn-doped Cd$_3$As$_2$ and shown that the band structure effects of Mn can be emulated with a $k\cdot p$ model for pristine Cd$_3$As$_2$ with large external applied fields. We found that an external field corresponding to a  Zeeman energy of 72 meV was needed to emulate the effects of just $2\%$ Mn-doping (ferromagnetically aligned). These Zeeman energies, and Mn doping by proxy, have substantial effects on the electronic structure, transforming the Fermi surface into nested ellipsoidal pockets. In the experimental reality, an applied external field is necessary to align the Mn moments at low doping. In this sense, the effects of an applied external field should be enhanced by the simultaneous magnetization of the Mn ions. 

Separate, nested ellipsoidal Fermi surfaces will each contribute uniquely to quantum oscillations. For fields applied along the a-axis as in the main text, let us assume the larger of the two nested ellipsoids has extremal area $S_\mathrm{Large}$ while the smaller ellipsoid has area $S_\mathrm{Small}$. We identify $f_\mathrm{Large}$ and $f_\mathrm{Small}$ as the corresponding quantum oscillation frequencies. In the main text, we have observed the main bulk quantum oscillation frequency of pristine Cd$_3$As$_2$ split into two peaks with low Mn-doping. We identify $f_\mathrm{Large}$ and $f_\mathrm{Small}$ from Fig. 4e in the main text. Using the theoretical framework we developed in Ref. \cite{Ness2025MnCd3As2}, we use $k_y-k_z$ slices of the band structure at various chemical potentials and Zeeman energies to theoretically calculate $S_\mathrm{Large}$ and $S_\mathrm{Small}$. We can make a correspondence between theory and experiment by noting that for some given chemical potential and Zeeman energy: 
\begin{equation*}
\frac{S_\mathrm{Large}}{S_\mathrm{Small}} = \frac{f_\mathrm{Large}}{f_\mathrm{Small}}
\end{equation*}

\begin{figure}
\includegraphics[width=0.98\linewidth]{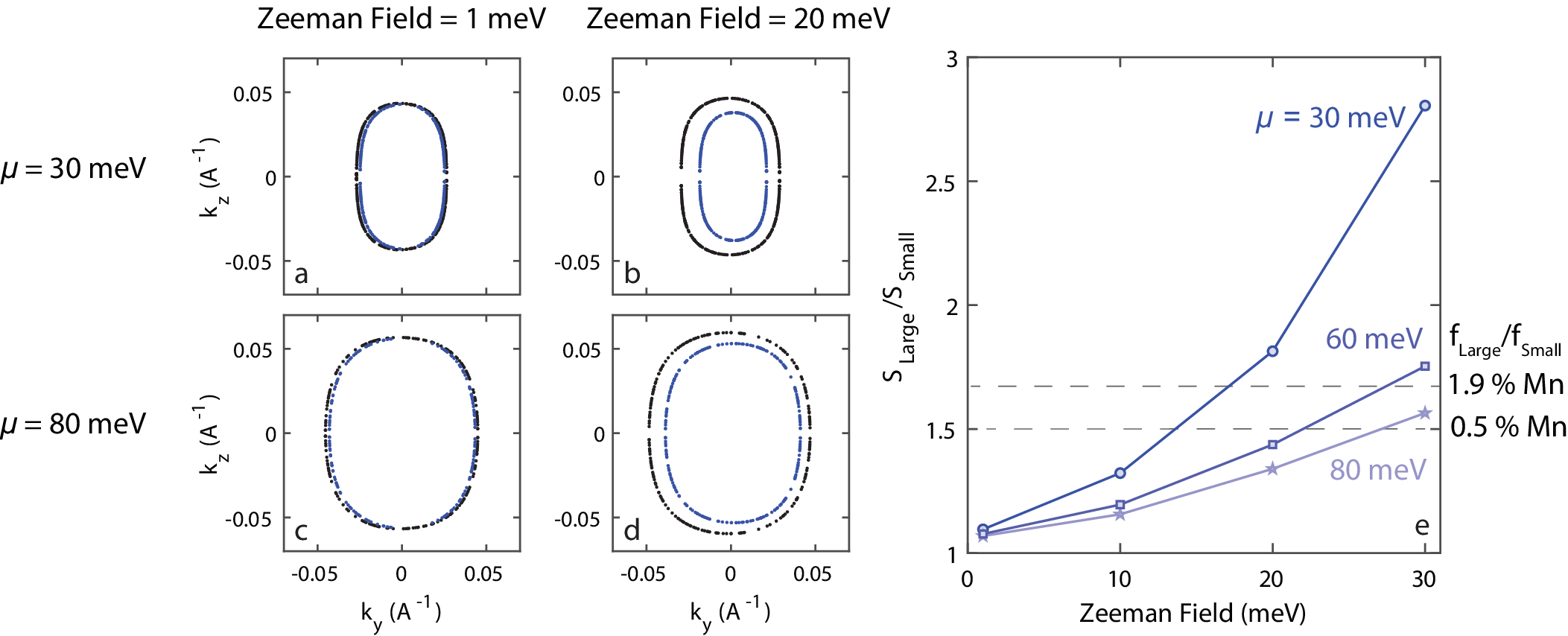}
\caption{\label{fig:SIBS} $k_y$-$k_z$ Band structure slices at $\mu =30$ meV (a,b) and $\mu=80$ meV (c,d). We have shown that large Zeeman energies caused by external fields can be used to simulate the effects of Mn on the band structure in Ref. \cite{Ness2025MnCd3As2}. Zeeman energies of 1 meV (a,c) and 20 meV (b,d) cause splitting of the Dirac point and alter the Fermi surface into nested ellipses. e) Plot of the ratio of the extremal area of the outer and inner ellipsoidal Fermi surfaces for fields along $k_x$ vs. Zeeman energy strength for various chemical potentials. Dashed lines represent the ratio of the split quantum oscillation frequencies discussed in the main text for 0.5$\%$ and 1.9$\%$ Mn doping. } 
\end{figure}

Figure \ref{fig:SIBS}a-d shows the effects of varying chemical potential and applied external magnetic fields (along $\hat{a}$), or ferromagnetically aligned Mn by proxy, on the electronic structure. Figs. \ref{fig:SIBS}a,b are calculated with $\mu=30$ meV  while Figs. \ref{fig:SIBS}c,d have $\mu = 80$ meV, corresponding to carrier densities of 7.1$\times10^{16}$ cm$^{-3}$ and 6.6$\times10^{17}$ cm$^{-3}$, respectively. The chemical potential is defined relative to the Dirac point. Figures \ref{fig:SIBS}a,c are calculated at Zeeman energies of 1 meV while Figs. \ref{fig:SIBS}b,d are calculated with a 20 meV Zeeman energy. In Fig. \ref{fig:SIBS}e, we use the $k_y-k_z$ slices to calculate the ratio of extremal areas for the ellipsoids, $S_\mathrm{Large}/S_\mathrm{Small}$, and plot the result against the Zeeman energy. Dashed horizontal lines represent the ratio $f_\mathrm{Large}/f_\mathrm{Small}$ calculated from the split quantum oscillation frequencies from Fig. 4e in the main text for 0.5$\%$ and 1.9$\%$. Using the DOS obtained from QSGW, we calculate the chemical potential for these samples as 60 meV and 67 meV, respectively. The Zeeman energy acts as a proxy to understand the effects of Mn on the band structure of Cd${}_3$As${}_2$. From the intersection of the experimentally calculated $f_\mathrm{Large}/f_\mathrm{Small}$ with the theoretically calculated $S_\mathrm{Large}/S_\mathrm{Small}$, we estimate the size of the Zeeman energy to be 22 and 29 meV for the 0.5$\%$ and 1.9$\%$ doped samples (as compared to the 72 meV Zeeman energy needed to theoretically emulate a 2$\%$ Mn doped sample where all moments are ferromagnetically aligned). Said another way, a Zeeman energy of 22(29) meV is needed to emulate the effects of 0.5$\%$ (1.9$\%$) Mn-doping on the pristine Cd${}_3$As${}_2$ band structure. Given the free moment of Mn$^{2+}$ is 5.92 $\mu_B$, the estimated Zeeman energy would correspond to  $1.8$-$2.4$ $\mu_B/\mathrm{Mn}$ (30-40$\%$ of the free Mn moment). This is consistent with the measured magnetization of the low doping samples in Fig. 3.

\providecommand{\noopsort}[1]{}\providecommand{\singleletter}[1]{#1}%